\documentclass{article}
\usepackage{emulateapj}
\usepackage{psfig}

\def\ltsima{$\; \buildrel < \over \sim \;$}
\def\gtsima{$\; \buildrel > \over \sim \;$}
\def\proptosima{$\; \buildrel \propto \over \sim \;$}
\def\simlt{\lower.5ex\hbox{\ltsima}}
\def\simgt{\lower.5ex\hbox{\gtsima}}
\def\simpropto{\lower.5ex\hbox{\proptosima}}

\makeatletter
\let\@internalcite\cite
\def\cite{\def\astroncite##1##2{##1\ ##2}\@internalcite}
\def\citey{\def\astroncite##1##2{##1\ (##2)}\@internalcite}
\def\@citex[#1]#2{\if@filesw\immediate\write\@auxout{\string\citation{#2}}\fi
  \def\@citea{}\@cite{\@for\@citeb:=#2\do
    {\@citea\def\@citea{; }\@ifundefined
       {b@\@citeb}{{\bf ??}\@warning
       {Citation `\@citeb' on page \thepage \space undefined}}%
{\csname b@\@citeb\endcsname}}}{#1}}
\def\@cite#1#2{#1\if@tempswa #2\fi}
\def\@biblabel#1{}
\def\astroncite#1#2{#1\ #2}
\makeatother

\begin{document}

\newcommand{\etal}{et al.}

\slugcomment{To be appeared in ApJ v554 n2 Jun 20, 2001 issue}

\lefthead{Y.Tsuboi et al.}

\righthead{X-rays from Class 0 in OMC-3}

\title{Discovery of X rays from Class 0 protostar candidates in OMC-3}

\author{Yohko Tsuboi}

\affil{Department of Astronomy \& Astrophysics, 525 Davey Laboratory, Pennsylvania State University, University Park PA 16802, USA}
\affil{tsuboi@astro.psu.edu}

\author{Katsuji Koyama\altaffilmark{1}, Kenji Hamaguchi}

\affil{Department of Physics, Faculty of Science, Kyoto University, Sakyo-ku, Kyoto 606-8502, Japan}
\affil{koyama@cr.scphys.kyoto-u.ac.jp, kenji@cr.scphys.kyoto-u.ac.jp}

\author{Ken'ichi Tatematsu, Yutaro Sekimoto}
\affil{Nobeyama Radio Observatory, National Astronomical Observatory of Japan, Nobeyama, Minamisaku, Nagano 384-1305, JAPAN}
\affil{tatematsu@nro.nao.ac.jp, sekimoto@nro.nao.ac.jp}

\author{John Bally, and Bo Reipurth}
\affil{Center for Astrophysics and Space Astronomy, University of Colorado, Boulder, CO 80309, USA}
\affil{bally@casa.colorado.edu, reipurth@casa.colorado.edu}

\altaffiltext{1}{CREST, Japan Science and Technology Corporation
(JST), 4-1-8 Honmachi, Kawaguchi, Saitama, 332-0012, Japan}

\begin{abstract}

We have observed the Orion Molecular Clouds 2 and 3 (OMC-2 and OMC-3)
with the Chandra X-ray Observatory (CXO).  The northern part of OMC-3
is found to be particularly rich in new X-ray features; four hard
X-ray sources are located in and along the filament of cloud cores.
Two sources coincide positionally with the sub$mm$-$mm$ dust
condensations of MMS 2 and 3 or an outflow radio source VLA 1, which
are in a very early phase of star formation. The X-ray spectra of
these sources show an absorption column of (1--3) $\times$10$^{23}$ H
cm$^{-2}$.  Assuming a moderate temperature plasma, the X-ray
luminosity in the 0.5--10 keV band is estimated to be $\sim$10$^{30}$
erg s$^{-1}$ at a distance of 450 pc. From the large absorption,
positional coincidence and moderate luminosity, we infer that the hard
X-rays are coming from very young stellar objects embedded in the
molecular cloud cores.

We found another hard X-ray source near the edge of the dust filament.
The extremely high absorption of 3 $\times$10$^{23}$ H cm$^{-2}$
indicates that the source must be surrounded by dense gas,
suggesting that it is either a YSO in an early accretion phase or a
Type II AGN (e.g. a Seyfert 2), although no counterpart is found at
any other wavelength.
 
In contrast to the hard X-ray sources, soft X-ray sources are found
spread around the dust filaments, most of which are identified with IR
sources in the T~Tauri phase.

\end{abstract}


\section{Introduction}

Young stellar objects (YSOs) evolve from a molecular cloud core
through protostellar and T Tauri phases to a main sequence star.
Protostars are generally associated with Class 0 and I energy
distributions, peaking in the {\it mm} and mid-to far infrared
wavelengths, respectively.  The Class I infrared objects appear to be
in a later phase of protostellar evolution.

One of the recent highlights in X-ray studies of YSOs is the discovery
of hard X-rays from embedded Class I objects, showing highly absorbed
spectra with occasional flares, or even quasi-periodic flaring
activity (Koyama et al. 1996, Grosso et al. 1997, Kamata et al. 1997,
Tsuboi et al. 2000). The overall X-ray phenomena are consistent with
enhanced solar-like magnetic activity, probably generated by a coupled
activity of convection and differential rotation on a star or between
a star and an inner disk (Tsuboi et al. 2000; Montmerle et al. 2000).
X-rays from protostars may be related to the competing processes of
accumulation of angular momentum towards the growing central star vs.
release of angular momentum by outflow processes. In this context,
X-ray observations of Class 0 objects in the dynamical infall phase
would be potentially important for the study of protostellar
evolution.
 
The infrared to {\it mm} radio spectrum of Class 0 objects can be
described by a black body model whose temperature is typically 15--30
K (e.g. Andr\'e, Ward-Thompson, \& Barsony 1993). Virtually all the
emission is from the surrounding envelope; no emission from a central
protostar is seen. Prominent outflows accompany such sources,
indicating disk accretion and release of angular momentum from the
accreting material. Class 0 protostars may have extremely large
absorption, up to 10$^{23-24}$ H cm$^{-2}$ or even more. Therefore
hard X-ray imaging is a unique technique to probe the central
protostars.

The Orion Molecular Cloud 2 and 3 regions (hereafter OMC-2 and OMC-3),
at a distance of about 450 pc (e.g. Genzel \& Stutzki 1989) exhibit a
chain of Class 0 objects detected at 1300 $\mu m$ (Chini et al. 1997),
350 $\mu m$ (Lis et al. 1998), and 450 and 850 $\mu m$ bands
(Johnstone \& Bally 1999). The size of the dust condensations is
typically 10--50 arcsec (FWHM) which corresponds to 5000--20000 AU
(FWHM) at 450 pc distance. The mass of each condensation is estimated
to be about 10 $M_\odot$, hence could be sites of low to intermediate
mass protostars (Chini et al. 1997). Evolution of protostars, from
Class 0 to Class I, can be traced from north to south in the clouds:
from mainly Class 0 sources in OMC-3 to mainly Class I sources in
OMC-2. Consequently, the OMC-2/3 region is a superb place to study
Class 0 and Class I sources with the same instrument, and
observationally trace the evolution of these protostellar phases.

	ASCA observed the OMC-2/3 region (Yamauchi et al. 1996), and
hard X-rays (2--10 keV band) are found from the densest regions of the
dust filament in the further analysis of the same data by Tsuboi
(1999). The hard X-ray distribution is well correlated with both the
ridge of molecular clouds and the cold dust condensations (1300 $\mu
m$ and 350 $\mu m$ peaks) within the limited spatial resolution of
ASCA. No excess in the soft X-ray band (0.7--2 keV) is found from the
OMC-3 region. Stimulated by this suggestion of hard X-ray emission
from inside the dust condensations (1300 $\mu m$ and 350 $\mu m$
peaks), and in order to find X-ray emission from Class 0 sources, we
have carried out Cycle 1 Chandra observations with more than 100 times
better spatial resolution than ASCA. This paper presents a first
report on the highlights of the new Chandra observations detecting
hard X-ray emission from the north of OMC-3, a site of the youngest
protostellar objects.

\section{Observations and Data Reduction}

	The OMC-2/3 observation was made in the New Year days of the
new millennium (2000 January 1--2) with the Chandra X-ray Observatory
(CXO) for 89.2 ks, having wide band sensitivity in the 0.2--10 keV
energy range with moderate energy resolving power and particularly
high spatial (sub-arcsec) resolution at the on-axis position. We used
the ACIS-I array consisting of four abutted X-ray CCDs, which covers
the complete OMC-2 and OMC-3 clouds, and one CCD from the ACIS-S array
covering a small region north of OMC-3.  In this paper we focus on and
discuss the data towards the northern part of OMC-3, which were
recorded with one of the ACIS-I CCDs.

We used Level 1 processed events provided by the pipeline processing
at the Chandra X-ray Center. To minimize the effect of the degradation
of charge transfer inefficiency (CTI), we applied the improved data
processing technique developed by Townsley et al. (2000).  To reject
background events, we applied a grade filter to keep only {\it ASCA}
grades\footnote{see
http;//asc.harvard.edu/udocs/docs/POG/MPOG/index.html} 0, 2, 3, 4, \&
6. We removed events from flaring pixels using the ``flagflare''
routine written by T.~Miyaji.  Artificial stripes caused probably by
hot pixels in the frame-store region and by particles which hit on the
CCD node boundaries were also removed. Detailed procedures for
cleaning low-quality events are given in {\it
http://www.astro.psu.edu/xray/axaf/recipes/clean.html}.

\section{Results}

\subsection{X-ray positions and identifications}

Figure 1 shows a 2$' \times 3'$ field image around the northern part
of OMC-3 in the 0.5--6 keV band overlaid with the 1300 $\mu m$
emission contour map from Chini et al. (1997). Red color represents
photon energies below 3 keV, while blue represents hard photons above
4 keV energy.  The absolute coordinates of the X-ray images are
fine-tuned to better than 0.1 arcsec, using the 2MASS infrared source
catalog 2MASS, of which details are given below.

For source finding, we used the {\it wavdetect} program in the Chandra
Interactive Analysis of Observations Software (CIAO\footnote{available
at http;//asc.harvard.edu/ciao/}, Version 1.0), which is based on a
Mexican hat wavelet decomposition and reconstruction of the image
(Freeman et al. 2000). We set the significance criterion at
1$\times$10$^{-5}$ and wavelet scales ranging from 1 to 16 pixels in
multiples of $\sqrt{2}$. The source detection was performed separately
in the soft (0.5--2 keV), hard (2--8 keV), and total (0.5--8 keV)
bands to improve sensitivity for non-embedded (soft X-ray band),
embedded (hard), and mildly embedded faint (total) sources,
respectively. In the 0.5--2 keV band, the contribution of the ASCA
grade 6 to the photons of point sources is about 1 \%, while that to
background is about 8 \% in the same band. In order to maximize the
signal-to-noise ratio, we hence used only the ASCA grades 0, 2, 3 and
4 for the source finding in the soft (0.5--2 keV) band, while the ASCA
grades 0, 2, 3, 4 and 6 are used in the other bands.

In the field shown in Figure 1, the source finding algorithm picked up
thirteen sources (No. 1--13 in Table 1) above a 4 $\sigma$
significance level in either of the three bands. We compared these
sources to the optical star catalog Hipparcos, and the infrared
catalog 2MASS. Only one counterpart is found from the optical catalog,
while two thirds of the sources have counterparts in the 2MASS
catalog. Selecting the five brightest X-ray sources, with statistical
errors within 0.3 arcsec at 90 \% confidence, we have made a
cross-correlation to the 2MASS infrared candidates, weighting by
photon counts. This resulted in a 1$''$.1 bore-sight correction to the
ACIS field; ACIS sources had offsets of $+$0$''$.51 and $-$0$''$.95
for right ascension and declination from the 2MASS frame,
respectively.  After the coordinate correction, the systematic
astrometric accuracy becomes better than 0.1 arcsec at 90 \%
confidence.  The thus corrected X-ray positions, counts in the 0.5--8
keV band, and hardness ratio are listed in Table 1, together with the
$J$, $H$, and $K$ magnitudes of the infrared counterparts and the
offset angle between the X-ray and IR positions. Since the offsets and
nominal absolute errors of the 2MASS sources are about 0.1--0.3 arcsec
and 0.1 arcsec, our X-ray position errors of sources No. 1--13 should
be 0.1--0.4 arcsec.

From figure 1, we see the hard X-ray sources more concentrated to the
dust emission in the 1300 $\mu m$ band (Chini et al. 1997). Among the
hard X-ray sources, two sources (No. 8 and 10 in Table 1) coincide
with positions of the dust condensations MMS 2 and MMS 3\footnote{The
coordinates of MMS 3 given in Table 1 of Chini et al. was incorrect
due to their problem in a Gaussian fit. Then we referred the correct
coordinate $\alpha = $5$^h$32$^m51.3^s$, $\delta =
-5^d02^m44.3^s$ in B1950 (Sievers A. private communication).}
(Chini et al. 1997) and CSO 6 and CSO 7 (Lis et al. 1998) within
position errors of $\sim5''$.

We further checked the hard X-ray enhancements from the other Class 0
protostar candidates or the dust condensations in the field of Figure 1.
The 2--8 keV band photon fluxes in a 5$''$ radius circle around the
1300 $\mu m$ (MMS 1--4, Chini et al. 1997) and 350 $\mu m$ (CSO 4--8,
Lis et al. 1998) sources are all below 3 $\sigma$ significance
level except for MMS 2 ($=$ CSO 6) and MMS 3 ($=$ CSO 7).

Since we found a complex structure around source No. 8, we made
closed-up views near this source in the 0.5--3 keV and 3--6 keV bands,
which are shown in figure 2. The position of 2MASS sources are
indicated by crosses. In the close vicinity of source No. 8, we can
clearly see one source in the 0.5--3 keV and two sources in the 3--6
keV band, although the source finding algorithm has missed them. We
here refer to these sources as $a$, $b$, and $c$, as indicated in
Figure 2. The X-ray positions are listed in Table 1.  Neither optical
nor IR counterparts are found towards the hard band sources $a$ and
$b$, but the soft band source $c$ coincides with a 2MASS source with
the $J$, $H$, and $K$ magnitudes of $>$16.9, $>$15.2, and 11.4,
respectively.

\begin{figure*}
\begin{minipage}[htbp]{0.9\textwidth}
  \begin{center}
  \mbox{\psfig{file=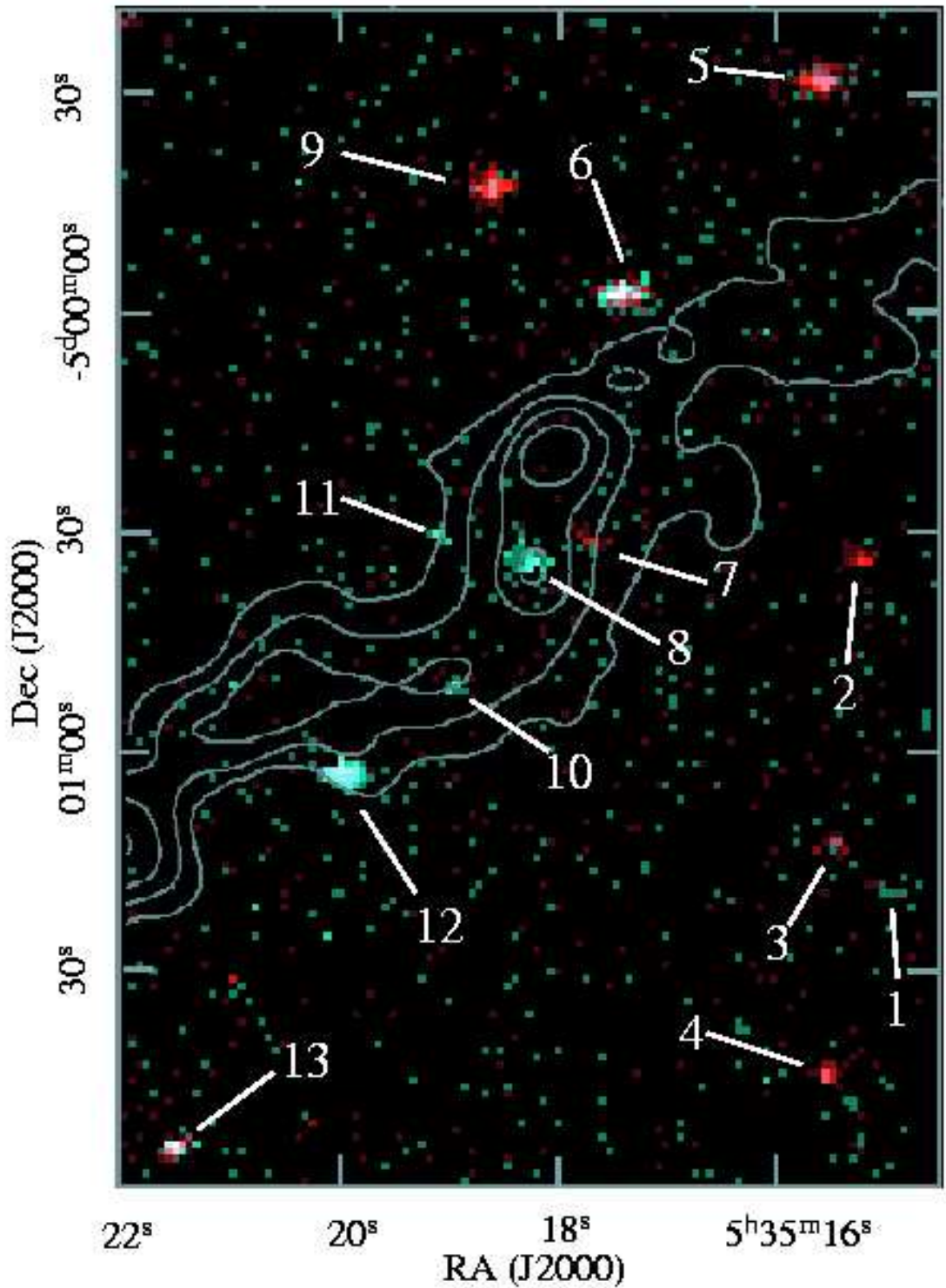,width=0.9\textwidth}}\\
  \end{center}
\end{minipage}~~
\figcaption{\small 
A 2$' \times 3'$ field image around the northern
part of OMC-3 in the 0.5--6 keV band overlaid with a 1300 $\mu m$
emission contour map from Chini et al. (1997). Red color represents
the photon energy below 3 keV, while blue represents hard photons
above 4 keV energy. The absolute coordinates of the X-ray images have
been fine-tuned to better than 0.1 arcsec, by reference to the 2MASS
infrared source catalog (see the text). Source numbers in table 1 are labeled.}
\end{figure*}
\vspace{0.25cm}

\begin{figure*}
\begin{minipage}[htbp]{0.9\textwidth}
  \begin{center}
  \mbox{\psfig{file=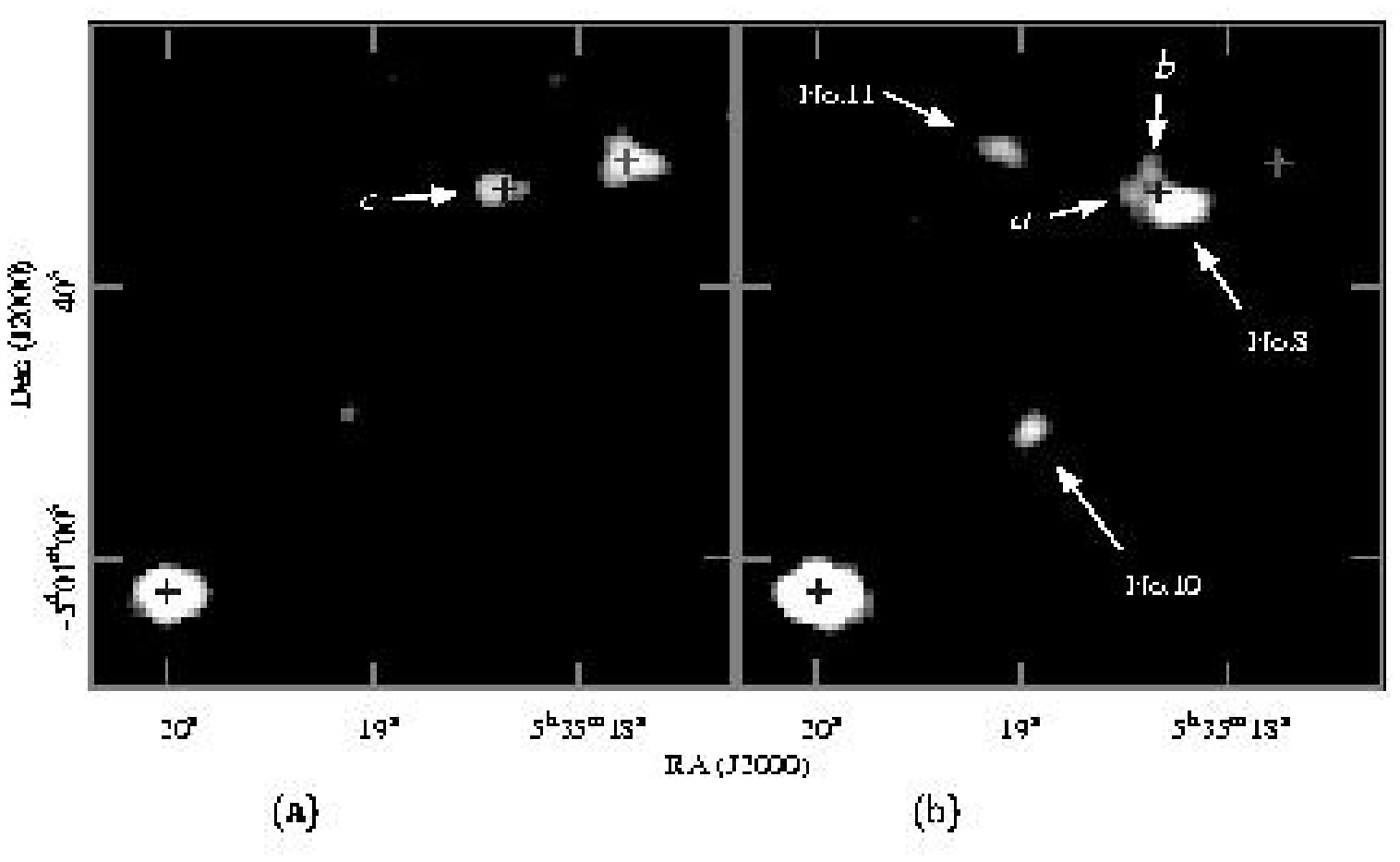,width=\textwidth}}\\
  \end{center}
\end{minipage}~~
\figcaption{\small 
Closed-up views near the source No. 8 in the 0.5--3
keV (right panel) and 3--6 keV bands (left panel). The position of
2MASS sources are indicated by crosses. Neither optical nor IR
counterparts are found at the hard band sources $a$ and $b$, but the
soft band source $c$ coincides with the position of a 2MASS source.}
\end{figure*}
\vspace{0.25cm}


\subsection{X-ray spectra}

\subsubsection{Bright Sources}

The brightest five sources in Figure 1 have reasonable statistics
(more than 100 photons), and we therefore made the X-ray spectra by
selecting 4$''$--9$''$ radius circles around the sources to include
$\sim$99 \% and $\sim$90 \% photons at 1.5 keV and 10 keV,
respectively.  The background spectra are made from an annulus with
inner and outer radii of 1.0 and 1.3 times of the source radius,
respectively.  We fitted the background subtracted spectra with a thin
thermal plasma model (MEKAL model; Mewe, Gronenschild, \& van den Oord
1985, Mewe, Lemen, \& van den Oord 1986, Liedahl, Osterheld, \&
Goldstein 1995) with interstellar absorption.  The effective area of
the telescope mirrors and the detection efficiency of ACIS were
calculated with the {\it mkarf} program in CIAO, Version 1.0.  In the
fitting, we fixed the abundances to 0.3 solar, referring to the
results for many YSOs in various star forming regions obtained from
the earlier ASCA observations (ex. Orion: Yamauchi et al. 1996, Rho
Oph: Kamata et al. 1997). The best fit values are listed in Table 2.

In order to justify our spectra fitting procedure, we selected the
ASCA target No. 7 in Yamauchi et al. (1996) and compared our Chandra
results.  This source lies 9 arc-min south-east from the center of
Figure 1. Both our result and that in Yamauchi et al. (1996) are in
reasonable agreement; the temperature, abundance, and the absorbing
column, are 2.7 vs. 2.2 keV, 0.19 vs. 0.20 solar, and 1.1 $\times$
10$^{21}$ vs. 0.8 $\times$ 10$^{21}$ H cm$^2$ (we referred to the
1-temperature model obtained with SIS and GIS simultaneous fitting in
Yamauchi et al. 1996). Therefore even with significant calibration
uncertainties in the early phase of the CXO observations, our fitting
gives reliable results.

\subsubsection{Hard X-ray Sources}

Although the statistics are limited, we examined spectra of the new
hard X-ray sources No. 8, 10 and 11. To minimize contamination from
the three sources ($a$, $b$ and $c$) located in the close vicinity of
source No. 8, we extracted the spectrum for No. 8 from a small circle
of 2 arcsec radius centered on No. 8. Then now, the contamination from
sources $a$ and $b$ to the region is negligible. As for No. 10 and 11,
we also used the same radius circles in order to treat the hard source
spectra uniformly. Using the software package
XSPEC\footnote{http://xspec.gsfc.nasa.gov/} and
MARX\footnote{http://spce.mit.edu/ASC/MARX} (version 3.01), we
calculated the effective area for this region smaller than the Chandra
Point Spread Function at the off-axis position of these sources.
These small regions include a background of at most 1 photon, hence we
made no background subtraction for these hard sources.

The spectra of No. 10 and 11 are fitted with a thin thermal plasma
model (MEKAL model) with 0.3 solar abundance and absorption. However,
essentially no constraints on the temperature and absorption are
obtained, since these two parameters are strongly coupled. We hence
fixed the temperature at two extremes, 1 and 5 keV, referring the best
fit temperature of the soft sources.  The absorptions are constrained
to be larger than 10$^{23}$ H cm$^{-2}$. We next assumed the
temperature to be 3.2 keV, because the hardest source (No. 12) among
the bright five has 3.2 keV plasma, and previous results towards very
young stars (Class I) are more or less around this value. The best-fit
spectra and parameters are shown in Figure 3 (b) and (c) and listed in
Table 2.

Figure 3 (a) indicates the spectrum obtained from the circle centered
on No. 8. It is more complicated than the others, showing a flux
minimum at about 3 keV, which implies the presence of two spectral
components: one is hard and heavily absorbed component (dashed line)
from No. 8 and the other is soft and less absorbed component that is
contamination from the source $c$ (dash-dotted line) separated from
No. 8 by 2$''$. Then we fitted the spectrum with two thin thermal
plasma (MEKAL model) components with independent absorptions. Since
the 2MASS counterpart of source $c$ has flux ratios in $J$, $H$, and
$K$ consistent with those of source No. 12, we assumed the X-ray
temperature and absorption for the contamination from $c$ (soft
component) to be the same as those of No. 12. The absorption of the
hard component is found to be larger than 10$^{23}$ H cm$^{-2}$, with
weak dependence on the assumed temperature (1 and 5 keV). In Table 2,
we give the best-fit model in which we assume the temperature of the
hard component to be 3.2 keV. The lower energy component is exactly
the same as that of spill-over photons from $c$ to the source region
of No. 8 (3 counts) which was predicted by simulation with XSPEC and
MARX.

\begin{figure*}
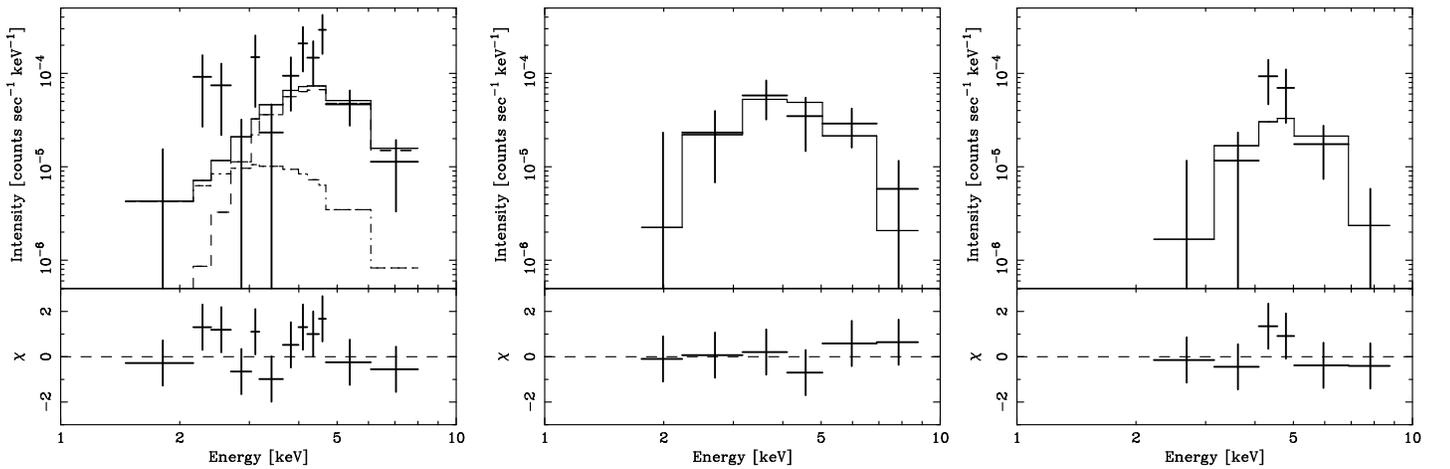

\begin{minipage}[htbp]{0.33\textwidth}
  \begin{center}
  \mbox{\psfig{file=fig3a.eps,angle=270,width=\textwidth}}\\
  \end{center}
\end{minipage}~~
\begin{minipage}[htbp]{0.33\textwidth}
  \begin{center}
  \mbox{\psfig{file=fig3b.eps,angle=270,width=\textwidth}}\\
  \end{center}
\end{minipage}
\begin{minipage}[htbp]{0.33\textwidth}
  \begin{center}
  \mbox{\psfig{file=fig3c.eps,angle=270,width=\textwidth}}\\
  \end{center}
\end{minipage}
\figcaption{\small 
The spectra extracted from a circle of 2 arcsec
radius centered on No. 8 (figure 3a), 10 (figure 3b) and 11 (figure
3c). The spectrum shown in figure 3a contains two components; one is
hard and heavily absorbed component (dashed line) from source No. 8 and
the other is soft and less absorbed component that is contamination
from source $c$ (dash-dotted line). Sources No. 10 and 11 are fitted
with a coronal plasma model (MEKAL model).  The lower panel shows the
residuals from the best-fit model.}
\end{figure*}
\vspace{0.25cm}

\subsection{Time Variability}

To obtain temporal variations of individual X-ray sources, we applied
a Kolmogorov-Smirnov test under the assumption of constant flux for
individual sources using the
XRONOS\footnote{http://xronos.gsfc.nasa.gov/} (Ver. 4.02) software
package. The results are listed in Table 1.  Among the sources, only
one source (source No. 9 in Table 1) exhibited a clear flare.

\section{Discussion}

We have detected about one dozen X-ray sources in the northern part of
OMC-3. This region has been observed in X rays with $Einstein$ (Ku \&
Chanan 1979, Vaiana et al. 1981, Caillault \& Zoonematkermani 1989,
Gagn\'e et al. 1994), with $ROSAT$ (Gagn\'e et al. 1995, Alcal\'a et
al. 1996) and with $ASCA$ (Yamauchi et al. 1996), however, no X-ray
sources has been detected in the previous observations. Then all the
X-ray sources in Table 1 are new X-ray objects. All the soft band
sources (hardness ratios in Table 1 are less than 0.2; No. 5, 6, 9,
12, and 13) have IR counterparts, most of which are likely to be
TTSs. The X-ray properties of the the soft band sources, i.e., plasma
temperatures (a few keV), luminosities (10$^{30}$ erg s$^{-1}$), and
time variabilities including the detection of a flare (for source
No. 9) are consistent with those of TTSs previously reported
(ex. Kamata et al. 1997).

The most notable result is the discovery of the highly absorbed X-ray
sources No. 8 and 10 coincident within the absolute position error of
$\sim5''$ with the dust condensations MMS 2 ($=$ CSO 6) and MMS 3 ($=$
CSO 7).  Since the background X-ray source density above the present
detection limit (10$^{-14}$ erg cm$^{-1}$ s$^{-1}$) is about 1 source
arcmin$^{-2}$, the probability to detect an X-ray source by chance
within the error circle of the millimeter sources (5 arcsec radius
circle) is about 2 \%. On the other hand, Reipurth, Rodr\'iguez, \&
Chini (1999) found a VLA source (VLA 1) towards MMS 2 with absolute
position accuracy of 1$''$.  It is only 1$''$.4 apart from the source
No. 8, hence these are certainly associated with each other.
Together with the large column of No. 8 and No. 10 of 1--3
$\times$10$^{23}$ H cm$^{-2}$, we therefore conclude that the hard
X-ray sources No. 8 and 10 are located in the dust condensations MMS 2
and 3.

The flux ratios of 350 $\mu m$ to 1300 $\mu m$ for MMS 2 ($=$ CSO 6)
and MMS 3 ($=$ CSO 7) are 99 and 87\footnote{The peak flux density of
MMS 3 in Table 1 in Chini et al. 1997 was incorrect. Then we referred
the correct flux 300 mJy (Sievers A. private communication).},
respectively, which are somewhat larger than the average of the other
MMS sources in OMC-3. However, no systematic trends are apparent. The
MMS 3 seems to be associated with a shock-excited H$_2$ flow D (Yu,
Bally, \& Devine 1997), plotting the correct coordinate (Sievers
A. private communication) on Figure 4 of Yu et al. (1997). MMS 2 is
associated with a prominent H$_2$ flow B (Yu, Bally, \& Devine 1997),
a Herbig-Haro object HH 331 (Reipurth 1999), molecular outflow
($^{12}$CO $J = 2 - 1$, Yu et al. 2000), and 3.6 cm emission (VLA 1,
Reipurth, Rodr\'iguez \& Chini 1999), which is most likely due to a
thermal jet. In addition, Aso et al. (2000) detected unresolved
HCO$^+$ outflow around MMS2--4 region. Given these powerful outflow
activities and considering that the two sources are parts of a chain
of similar sources of which at least six are documented Class 0
sources, MMS 2 and 3 are most probably also Class 0 sources. Besides
them, the absorption columns of MMS 2 and 3 obtained by our
observation ($>10^{23}$ cm$^{-2}$) are one order of magnitudes larger
than those of Class I protostars obtained with ASCA (Kamata et
al. 1997, Tsuboi et al. 2000) and Chandra (Imanishi, Koyama \& Tsuboi
2001), which highly supports that they are Class 0 sources. Therefore
it is likely that this is the first discovery of X-rays from YSOs
preceding the Class I phase.

No significant X-ray enhancement is found from other dust
condensations than MMS 2 and 3 in the northern part of OMC-3. One may
attribute it to much more larger columns (larger mass) than for MMS 2
and 3, but simple estimates based on the mass and size suggest that
the column densities of them are 10$^{23}$--10$^{24}$ H cm$^{-2}$, and
no systematic difference from MMS 2 and 3 are seen.

The soft source $c$ is probably unrelated to the MMS 2 condensation,
and may be located in front.  However, the hard sources $a$ and $b$
are likely to be located within the MMS 2 core. If this is so, the MMS
2 core is a site of multiple star formation.

Source No. 11, although located at the edge of the dust filament,
exhibits a large absorption of 3$\times$10$^{23}$ H cm$^{-2}$, which
exceeds the expected column of the cloud at this position. Thus the
large absorption should be very local to the source No. 11, possibly
either infalling gas around a protostar (protostellar scenario) or a
torus surrounding an active galactic nucleus (type II AGN scenario).

We express our thanks to the High-energy Astrophysics Group at the
Penn State University led by Prof. G. Garmire. Most of the data
reduction procedure and analysis were made using the software packages
and methods developed by this team. We also gratefully acknowledge the
assistance of Prof. E. D. Feigelson, Dr. Y. Maeda, and
Dr. L. Townsley. We are grateful to Dr. A. Sievers for information
about 1300 $mu m$ emission from MMS 3. A part of this work is
supported by the JPS grant of collaboration with foreign country
(grant No. 10147103). YT is financially supported by JSPS.

\normalsize


{\bf Reference}\\

Andre, P., Ward-Thompson, D. \& Barsony, M. 1993, ApJ 406, 122

Aso, Y., Tatematsu, K., Sekimoto, Y., Nakano, T., Umemoto, T., Koyama, K., Yamamoto, S. 2000, ApJS in press

Caillault, J-P \& Zoonematkermani, S. 1989, ApJL 338, L57

Chini, R., Reipurth, Bo., Ward-Thompson, D., Bally, J., Nyman, L-A, Sievers, A., \& Billawala, Y., 1997, ApJ 474, L135

Freeman, P. E., Kashyap, V., Rosner, R., \& Lamb, D. Q. 2000, ApJ, submitted

Gagne, M. \& Caillault, J-P. 1994, ApJ 437, 361

Gagne, M., Caillault, J-P., \& Stauffer, J. R., 1995 ApJ 445, 280

Genzel, R. \& Stutzki, J. 1989, ARA\&A 27, 41

Grosso, N., Montmerle, T., Feigelson, E. D., Andre, P., Casanova, S., \& Gregorio-Hetem, J. 1997, Nature 387, 56

Imanishi, K., Koyama, K., \& Tsuboi, Y. 2001 in preparation

Johnstone, D. \&  Bally, J. 1999, ApJL 510, L49

Kamata, Y., Koyama, K., Tsuboi, Y., \& Yamauchi, S. 1997, PASJ 49, 461

Koyama, K., Hamaguchi, K., Ueno, S., Kobayashi, N., \& Feigelson. E. 1996, PASJ 48, L87

Ku W. H.-M. \& Chanan, G. A. 1979, ApJL 234, L59

Liedahl, D. A.,  Osterheld, A. L., Goldstein, W. H. ApJL 438, 115

Lis, D. C., Serabyn, E., Keene, Jocelyn, Dowell, C. D., Benford, D. J., Phillips, T. G., Hunter, T. R., \& Wang, N. 1998, ApJ 509, 299

Mewe, R., Gronenschild, E. H. B. M., \& van den Oord, G. H. J. 1985, A\&AS 62, 197

Mewe, R., Lemen, J. R, \& van den Oord, G. H. J. 1986, A\&AS 65, 511

Montmerle, T., Grosso, N., Tsuboi, Y., \& Koyama, K. 2000, ApJ 532, 1089

Reipurth, B. 1999, {\em A General Catalogue of Herbig-Haro Objects},
2. Edition, available at http://casa.colorado.edu/hhcat

Reipurth, B., Rodr\'iguez, L. F., \& Chini, R. 1999, ApJ 118, 983

Townsley, L. K., Broos, P. S., Garmire, G. P., \& Nousek, J. A. 2000, ApJL 534, L139

Tsuboi, Y., Imanishi, K., Koyama, K., Grosso, N., \& Montmerle, T. 2000, ApJ 532, 1097

Tsuboi, Y. 1999, PhD thesis, see also ISAS RN 689

Vaiana, G. S., Fabbiano, G., Giacconi, R., Golub, L., Gorenstein, P., Harnden, F. R. Jr., Cassinelli, J. P., Haisch, B. M., Johnson, H. M.,  Linsky, J. L., Maxson, C. W., Mewe, R., Rosner, R., Seward, F., Topka, K., \& Zwaan, C. 1981, ApJ 245, 163

Yamauchi, S., Koyama, K., Sakano, M., \& Okada, K. 1996, PASJ 48, 719

Yu, K. C., Bally, J., \& Devine, D. 1997, ApJL 485 L45

Yu, K. C., Billawala, Y., Smith, M. D., Bally, J., Butner, H. M. 2000, AJ in press







\end{document}